\documentclass[%
 reprint,
superscriptaddress,
 amsmath,amssymb,
 aps,
]{revtex4-2}

\usepackage{mathtools}
\usepackage{graphicx}
\usepackage{xcolor}
\usepackage{enumerate}
\usepackage{wrapfig}
\usepackage{lipsum}  
\usepackage{dcolumn}
\usepackage{braket}
\usepackage{appendix}
\usepackage[colorlinks=True,citecolor={blue},linkcolor={blue},urlcolor={blue}]{hyperref} 

\usepackage{amsmath,amsthm,amssymb,amsfonts}
\usepackage{bm}

\begin{document}
\preprint{APS/123-QED}

\title{Nonlinear Quantum Optics at a Topological Interface Enabled by Defect Engineering}

\author{L. Hallacy}
\affiliation{Department of Physics and Astronomy, University of Sheffield, Sheffield S3 7RH, UK}%

\author{N.J. Martin}
\email{n.j.martin@sheffield.ac.uk}
\affiliation{Department of Physics and Astronomy, University of Sheffield, Sheffield S3 7RH, UK}%

\author{M. Jalali Mehrabad}
\affiliation{Joint Quantum Institute, University of Maryland, College Park, MD 20742, USA}%

\author{D. Hallett}
\author{X. Chen}
\author{R. Dost}
\author{A. Foster}
\author{L. Brunswick}
\author{A. Fenzl}
\affiliation{Department of Physics and Astronomy, University of Sheffield, Sheffield S3 7RH, UK}%

\author{E. Clarke}
\author{P.K. Patil}
\affiliation{EPSRC National Epitaxy Facility, University of Sheffield, Sheffield S1 4DE, UK}%


\author{A.M Fox}
\author{M.S.~Skolnick}
\author{L.R. Wilson}
\affiliation{Department of Physics and Astronomy, University of Sheffield, Sheffield S3 7RH, UK}%

\begin{abstract} 
The integration of topology into photonics has generated a new design framework for constructing robust and unidirectional waveguides, which are not feasible with traditional photonic devices. Here, we overcome current barriers to the successful integration of quantum emitters such as quantum dots (QDs) into valley-Hall (VH) topological waveguides, utilising photonic defects at the topological interface to stabilise the local charge environment and inverse design for efficient topological-conventional mode conversion. By incorporating QDs within defects of VH-photonic crystals, we demonstrate the first instances of single-photon resonant fluorescence and resonant transmission spectroscopy of a quantum emitter at a topological waveguide interface. Our results bring together topological photonics with optical nonlinear effects at the single-photon level, offering a new avenue to investigate the interaction between topology and quantum nonlinear systems.
\end{abstract}

\maketitle 
\begin{figure}[t]
    \centering
    \includegraphics[width=0.48\textwidth]{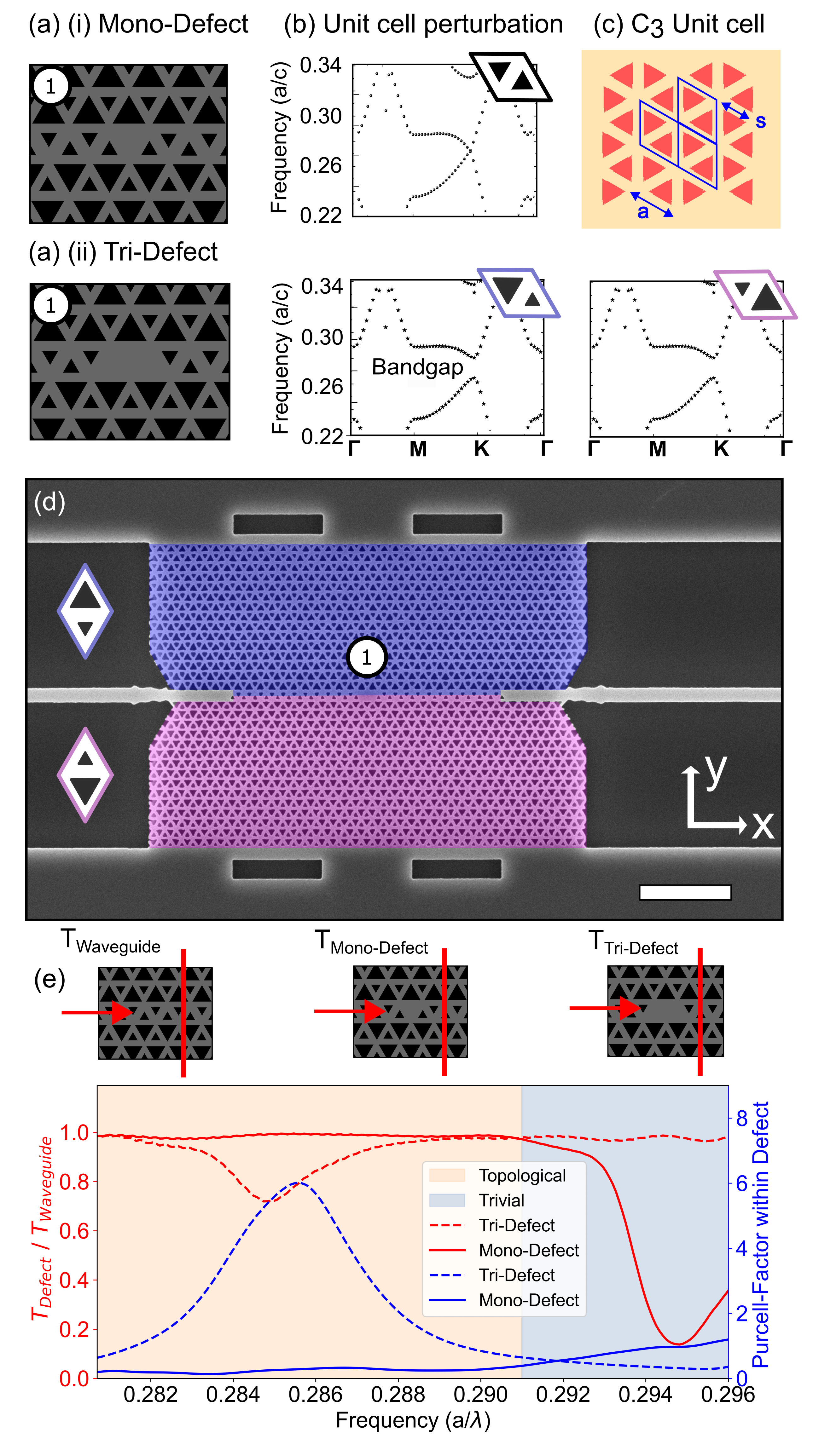}
    \caption{(a) Schematic diagram of the (i) single aperture removed (Mono) defect waveguide and (ii) triple aperture removed (Tri) waveguide-coupled defect cavity. (b) An illustration of the perturbed unit cells that form the VH-waveguide and their bandstrutures, showing the formation of a band gap at the K point. (c) The $C_{3}$ unit cell. (d) SEM image of the waveguide structure showing the location of the defect at point (1). (e) The frequency dependence of the transmission through the Mono and Tri Defects presented on the left Y-axis (red) with the Purcell enhancement at the centre of the defects presented on the right Y-axis (Blue).}
    \label{fig:intro}
\end{figure}
 \begin{figure*}[t]
    \centering
    \includegraphics[width=0.8\textwidth]{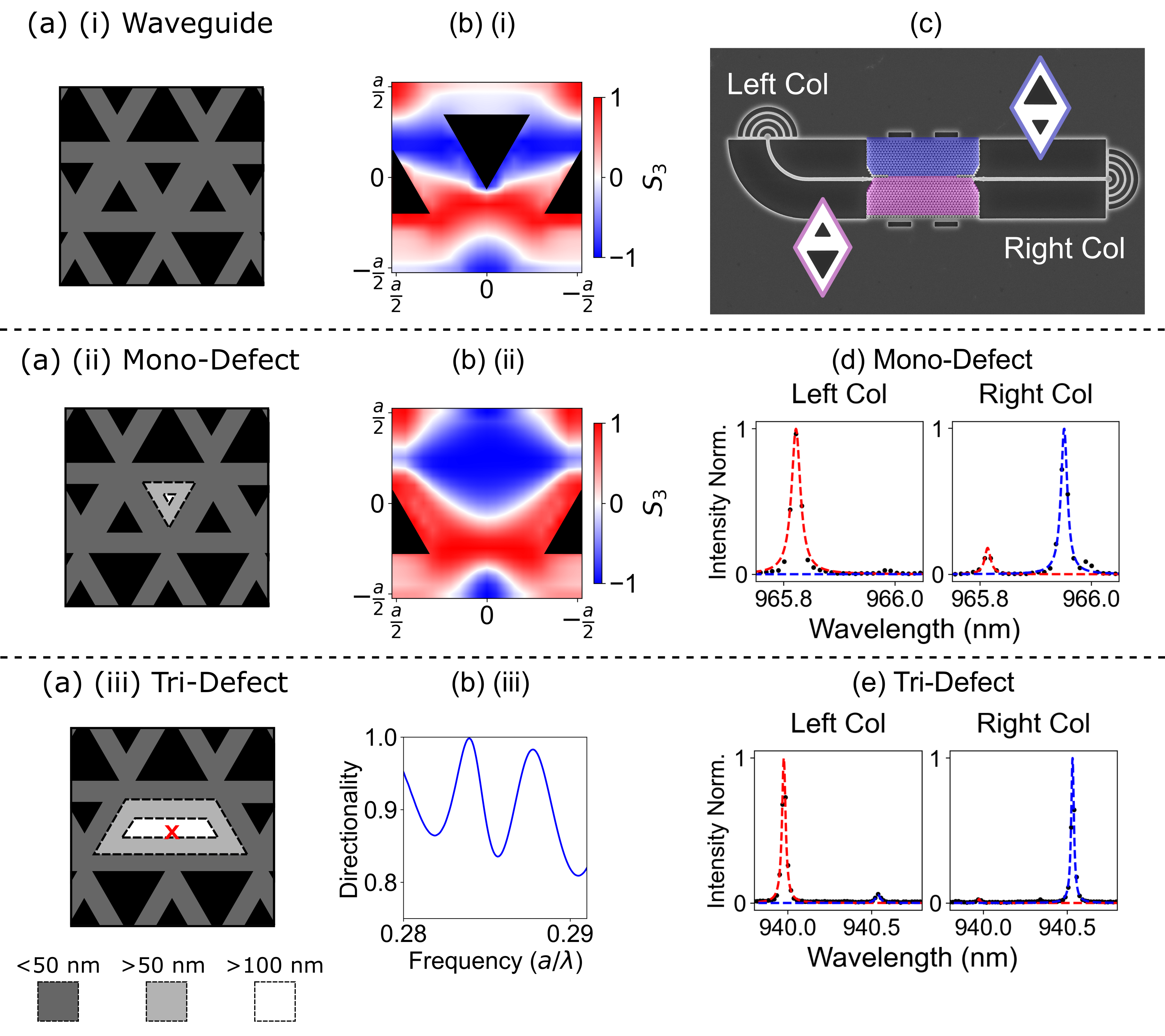}
    \caption{ Within (a) (i-iii), regions less than 50nm, greater than 50nm and greater than 100nm away from an etched surface are highlighted. (b) (i-ii) Shows the Stokes $S_{3}$ parameter for the localised electric field at the centre of the slab waveguide slab for the (i) un-modified waveguide and (ii) the Mono-defect waveguide. (b) (iii) Shows the equivalent frequency dependence of the directionality of an emitter placed within the Tri-defect cavity at the red cross indicated in (a)(iii). (c) SEM of the device showing the left and right collection ports (Col). (d) An example of high chiral contrast (85\%) PL emission, from a QD located within the Mono-defect. (e) An example of high chiral contrast (92\%) PL emission, from a QD located within the Tri-defect. }
    \label{fig:chiral}
\end{figure*}
\section{Introduction}



The last two decades have seen the emergence of topological photonics as a new and powerful approach for the design of photonic devices with novel functionalities \cite{LingLu2014, ozawa2019topological, price2022roadmap,zhang2023second,smirnova2020nonlinear,jalali2023topological,Khanikaev2024}. Many of these developments have been motivated by the fact that topological systems exhibit chiral or helical edge states that are confined to the boundary of the system and are remarkably robust against imperfections common to integrated photonic devices \cite{Wang2009, Rechtsman2013,Hafezi2013}. Examples of implementation of topological photonics in the linear regime include robust optical delay lines \cite{mittal2014topologically}, slow light engineering \cite{guglielmon2019broadband}, waveguides, tapers, and re-configurable routers \cite{shalaev2019robust,Flower2023,zhao2019non,zheng2024dynamic,jalali2024strain,sharp2024near}. While early efforts in topological photonics focused on linear devices, more recent demonstrations have included nonlinear effects, extending the scope of possible applications to include lasers \cite{St-Jean2017, Bahari2017, Bandres2018, Yang2022}, parametric amplifiers \cite{peano2016topological,sohn2022topological}, quantum light sources \cite{mittal2018topological, Blanco-Redondo2018, Mittal2021, Dai2022} and frequency combs \cite{flower2024observation}. A more recent and intriguing direction has been exploring strong light-matter coupling to induce strong interaction between photons. To achieve this, microcavity exciton-polaritons \cite{Klembt2018,dikopoltsev2021topological}, transition metal dichalcogenides \cite{Li2021}, and quantum dots (QDs) \cite{Barik666} were integrated in topological photonic devices. In particular, due to their scalability and high optical quality for on-chip single photon generation, there has been great interest in implementing QDs in topological photonic devices. For example, QDs have been utilised in various applications, such as internal light sources in topological photonic ring resonators \cite{JalaliMehrabad_APL}, topological 1D cavity lasers \cite{ota2018topological}, and topological slow-light waveguides \cite{yamaguchi2019gaas,Yoshimi:20}.They have also been used as integrated
single-photon emitters for cavity-QED in topological
nanocavities \cite{rao2024single,xie2020cavity}, as well as for chiral quantum optics in fast and slow-light topological waveguides \cite{barik2018topological,JalaliMehrabad_Optica,arakawa_single_photon_valley}, topological all-pass \cite{Barik_2020,JalaliMehrabad_Optica} and add-drop filters \cite{mehrabad2023chiral}.



Efficient integration of several QDs in topological photonics could lead to the realization of collective effects such as chiral super and sub-radiance effects and spin chains \cite{Lodahl2017}. However, development in this direction has remained elusive due to several remaining challenges. On one hand, position-dependence of the emitter's coupling efficiency and chiral coupling in QD-coupled topological waveguides is a significant limitation in these quantum optics interfaces as recently explored both theoretically and experimentally \cite{hauff2022chiral,martin2023topological}. Moreover, regions of high directionality and high coupling efficiency areas are mostly present in the holes of crystal rather than the material \cite{Nussbaum2022,JalaliMehrabad_Optica,martin2023topological}, which is detrimental for coupling to solid-state quantum emitters. Another remaining challenge is the efficiency of mode conversion at the topological-conventional waveguide interfaces \cite{JalaliMehrabad_Optica,Shalaev2018,Mehrabad2023}. Further optimization of such mode conversion is essential for the efficient integration of these optical components for scalable photonic circuitry. 

In this work, we report a novel platform based on the integration of QDs in lattice defects that overcomes many of the aforementioned challenges. We explore topological photonic crystals with defects illustrated in Fig. \ref{fig:intro} (a) (i-ii),(d) which aims to break the translational symmetry of the VH waveguide to provide an excellent region for high-efficiency integration of QDs with minimal impact on transmission through the topologically non-trivial operational regime of the waveguide. We demonstrate the defects can be utilised to achieve highly directional emission and we harness the improvements to the QDs local environment to achieve the first demonstration of resonance spectroscopy and
optical non-linear response of quantum emitter at a topological edge state. Additionally, we use an inverse design approach to improve the mode conversation efficiency of topological and non-topological regions (see supplementary section S1). 

\section{Device Design}

\subsection{Creation of a Topological Interface}

The VH topological photonic crystal (PhC) used in this work is created from a honeycomb lattice of triangular holes in a semiconductor membrane. For these semiconductor-based QD systems, we grow a GaAs-based p-i-n diode using molecular beam epitaxy with InAs Stranski-Krastanov dots positioned in the middle of the intrinsic layer. This wafer design (detailed in supplementary section S4) allows for fast tuning of embedded QDs via the quantum-confined Stark effect and reduces charge noise by modulating the electric field. The rhombic unit cell of the PhC contains a pair of triangular holes. Initially, with holes of equivalent diameters, the band structure of the PhC for TE polarization shows a Dirac cone at the \(K\) point (and similarly at the \(K'\) point), as shown in Fig. \ref{fig:intro}(b). By shrinking one triangle and expanding the other, a modified PhC supports a bandgap for TE polarized light. A notable aspect of the band structure is the opposite sign of the Berry curvature at the \(K\) and \(K'\) points, as demonstrated in \cite{He2019}. By interfacing two of these perturbed photonic crystal structures together, one the inverse of the other, a topological waveguide interface can be created. The difference in Berry curvature at the connection of the two PhCs results in the confinement of counter-propagating edge states with opposing helicity at the interface \cite{He2019}. The design used in this work utilizes a small triangle side length of \(L_{\rm S} = 0.7a/\sqrt{3}\) and a large triangle side length of \(L_{\rm L} = 1.3a/\sqrt{3}\), which establishes a single-mode, topologically non-trivial, slow light region within the waveguide \cite{Yoshimi:21}.

\subsection{Defect Engineering and QD Environment}

By optimizing solely around optical properties, we can inadvertently introduce challenges related to the environment of quantum emitters. Integrating QDs within stable environments is essential for high-performance single-photon emission and operation as few photon nonlinearities. Achieving this in VH waveguides requires careful optimization of the waveguide design, taking into account the stability of the environment of embedded quantum emitters. 

In the early studies of resonance fluorescence (RF) in semiconductor epitaxial QDs, researchers encountered several obstacles. The primary issue was the interaction of QDs with their surrounding charge environment, leading to phenomena such as spectral diffusion and linewidth broadening \cite{doi:10.1021/acsphotonics.0c00758}. These effects were primarily due to fluctuations in the local charge environment, including charge traps and defect states, which induced instability in the quantum dot emissions \cite{PhysRevLett.116.213601}. Additionally, non-radiative recombination processes, often exacerbated by these charge fluctuations, significantly quenched the emission, making it challenging to isolate and study individual excitonic states \cite{zhai2020low}. Over time, improvements in diode characteristics, fabrication, and weak non-resonant optical gating have minimized these issues in nanobeam and PhC-based structures \cite{nguyen2012optically, hallett2018electrical, doi:10.1021/acsphotonics.0c00758}. However, these methods alone have not been sufficient to observe RF in topological waveguides, such as VH waveguides. Observing stable charge states via RF in topological waveguides presents particular challenges. The main difficulty lies in the geometry required to observe broad, non-trivial topological modes within GaAs-based photonic structures. Specifically, a large amount of the wafer membrane is etched away, and the sections remaining are exposed to increased charge noise from charge traps at the etched surfaces \cite{PhysRevApplied.9.064019}. These charge traps degrade QD performance if they are less than 40 nm away from these surfaces \cite{chiral_stats, PhysRevApplied.9.064019}, resulting in higher spectral wandering. No point within the original VH waveguide is more than 50nm away from an etched surface Given these considerations, it is unsurprising that topological structures with embedded emitters encounter difficulties in achieving RF and emitter based, strong nonlinear interactions, significantly hindering their potential for scalable quantum systems with high coherence and indistinguishability.

To address these challenges, we remove etched holes within the unmodified topological waveguide (all illustrated in Fig \ref{fig:intro} (a) (i-ii) to create defect regions where QDs can be located without being affected by being in close proximity to etched surfaces, in regions that we refer to as Mono/Tri-defects respectively. Fig \ref{fig:intro} (e)) shows how the Mono-defect has little impact on the transmission through the topologically non-trivial spectral region of the waveguide's operation. The Tri-defect offers a significantly greater simulated maximum Purcell enhancement in comparison to the Mono-defect, with the trade off of a moderate decrease in the transmission. Further optimisation of these defects could lead to greater enhancements and reduced transmission loss, as has been developed for conventional photonic crystal systems \cite{ChiralCav_2022}.

We believe that this method significantly reduces the adverse effects of charge instability caused by surface states and does not compromise the coupling efficiency of QDs into non-trivial topological modes. Additionally, we use inverse design techniques to create an efficient mode converter between a slab waveguide and a non-trivial topological mode to improve the system's scalability (see supplementary section S1). Such an approach promises to leverage the unique advantages of topological waveguides while circumventing the limitations imposed by conventional topological PhCs.

\section{Results}

\begin{figure*}[t!]
    \centering
    \includegraphics[width=0.9\textwidth]{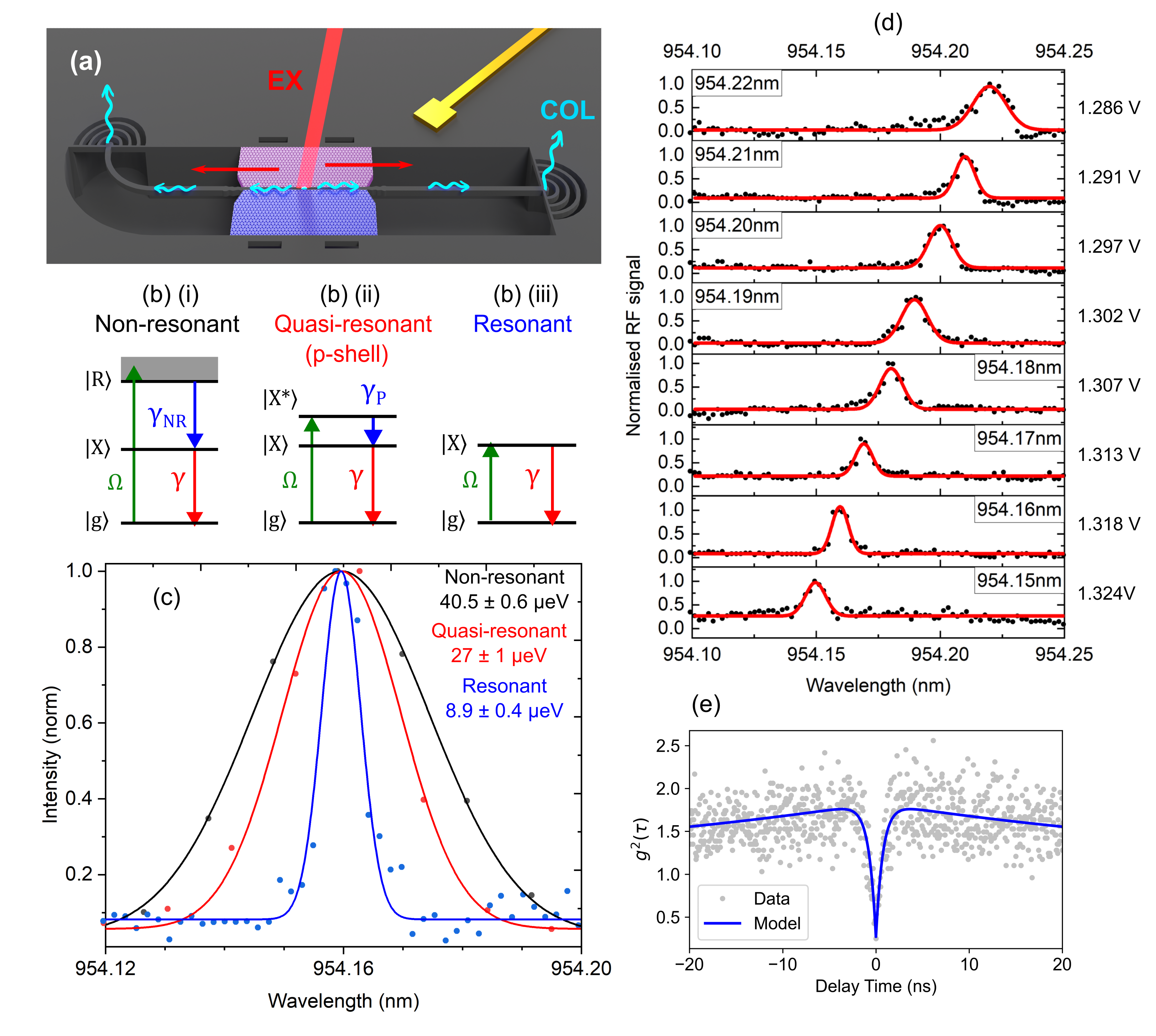}
    \caption{ (a) Render of topological PhC with Tri-defect defect showing excitation scheme for resonant measurements. (b) (i) Energy level diagram illustrating non-resonant driving from ground state $\ket{g}$ (at energy $\Omega$) to above band state $\ket{R}$ and recombination to intermediate charge state $\ket{X}$ with non-resonant radiative emission energy $\gamma_{NR}$ and back to ground state with resonant emission energy $\gamma$. (ii) Energy level diagram for the quasi-resonant scheme with P shell state $\ket{X^*}$ (determined from the difference in pumping and emission wavelength $\Delta \lambda = 14.83nm $) and intermediate emission $\gamma_{p}$ to charge state  $\ket{X}$. (iii) Energy level diagram for resonant scheme directly exciting charged excitonic state. (c) Comparison of linewidth in non-resonant, quasi-resonant, and resonant excitation showing a reduction in consistent reduction in spectral wandering (Gaussian fit of linewidth given in top right corner) .Non-resonant and quasi-resonant spectra obtained via CCD spectrometer scans while resonant spectra is obtained from bias modulated photon-intensity with background subtraction (further details in supplementary section S6) while scanning laser with wavelengths shown on x-axis in plot. (d) RF from a QD in a topological defect waveguide. RF signal obtained by scanning bias and measuring APD flux and normalised. Minimum linewidth observed was $8.9 \pm  0.4\mu eV$ at 954.16nm. (e) Autocorrelation measurement of single dot emission under quasi-resonant excitation.}
    \label{fig:resonance}
\end{figure*}

\begin{figure*}[t]
    \centering
    \includegraphics[width=0.99\textwidth]{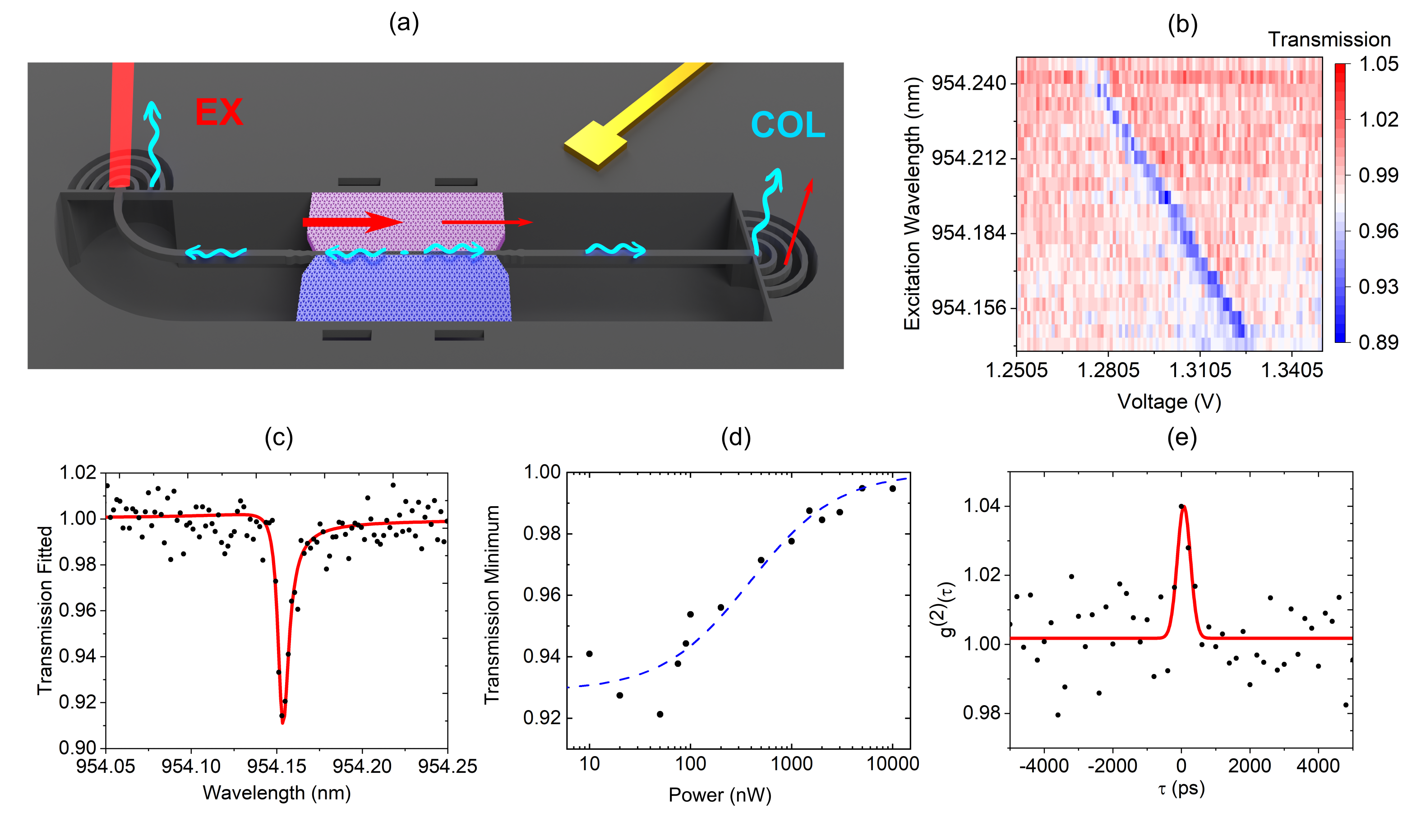}
    \caption{(a) Diagram showing resonant scattering in the Tri-defect transmission. The Render shows the resonant transmission scheme, where the pump laser and far-field collection are at separate out-couplers.(b) Resonant transmission scan measured by sweeping voltage under a resonantly driven CW laser (at 50nW). (c) An example of transmission spectra observed at 50nW power at $\lambda_L = 954.16$nm with a fitted maximum dip of ~8\%with a fano-lineshape. (d) Normalized transmission on resonance with the spectral line as a function of laser power, for $\lambda_L$ = 954.16nm. (e) Second-order autocorrelation function for photons transmitted through the topological waveguide and defect region at zero detuning. We observe clear bunching at $\tau$ = 0 (peaking at 1.038) using 200ps bin width. }
    \label{fig:transmission}
\end{figure*}
\subsection{Observation of High Directional Contrast}

A documented challenge for topological waveguides \cite{martin2023topological, Nussbaum2022} has been reliably creating highly directional emission. Here we demonstrate that the introduction of the Mono and Tri-defects are a way to create highly directional QD emission. Fig \ref{fig:chiral} (b) (i-ii) show the Stokes $S_{3}$ parameter, which for the un-modified waveguide and Mono-defect characterises the degree of circular polarisation of the localised electric field. The strong light confinement in these structures locks the local polarization of the light to its propagation direction. This interplay between polarization and propagation direction leverages spin-orbit coupling, where the spin state associated with a particular transition of a quantum emitter determines its polarization and thus its emission direction \cite{Lodahl2017}. For the un-modified waveguide, regions of the waveguide that can achieve high directionality are located close to etched surfaces ($<$50nm) from the waveguides apertures. Removing one of these apertures to form the Mono-defect greatly increases the distance from the peak of the $S_{3}$ value to any etched surfaces, and creates a better spatial overlap of the electric field intensity and the $S_{3}$ map (see supplementary section S3) . This distance is even greater for the Tri-defect. In the case of the Tri-defect, the mechanism for the directionality arises from a interference effect between the two cavity modes it supports \cite{ChiralCav_2022} with the wavelength dependence of the directionality shown in Fig \ref{fig:chiral} (b) (iii). More information on this mechanism for directional coupling within the Tri-defect cavity can be found within supplementary section S2. 

In order to demonstrate directional emission from QDs in these defect structures, $\mu$PL measurements of individual QDs were carried out, measuring from the left and right ports of the waveguide, as shown in the SEM of Fig \ref{fig:chiral} (c). The resulting photoluminescence (PL) spectra from a single representative QD under a magnetic field in the Faraday geometry of $~$2T are illustrated in Fig \ref{fig:chiral} (d) and (e) for the Mono-defect and Tri-defect respectively. These spectra reveal two Zeeman-split states, exhibiting an asymmetric intensity for the $\sigma^{+}$ and $\sigma^{-}$ polarized transitions. Notably, the intensity asymmetry reverses when collecting PL from the opposite optical collection direction, indicating directional emission. The measured directional contrast for these two examples, reached a maximum of 85$\%$ for the Mono-defect and 92$\%$ for the Tri-defect. It is common to observe asymmetry in the contrast measured in either direction in experiments of this nature \cite{Mehrabad2023,Cole2017}, an effect that is visible in these results, and not yet fully understood.


\subsection{Quasi-Resonant and Resonant Excitation}

To measure RFfrom a QD within the Tri-defect, we use the excitation scheme and device design illustrated in Fig \ref{fig:resonance} (a). This consists of a VH waveguide with a Tri-defect placed in the centre of the waveguide. The PhC is coupled via the inversely designed mode converter to two deep-etched Bragg grating couplers for far field collection.

Fig \ref{fig:resonance} (b) (i-iii) compares the energy level diagrams of non-resonant, quasi-resonant and resonant excitation schemes of a QD. In above-band excitation (b)(i), photons with energy greater than the bandgap of the QD material are used to create electron-hole pairs in the bulk material, which then relax through phonon emission and are captured by the quantum dot. The multiple steps and environmental interactions involved in this process lead to dephasing and broader emission linewidths compared to more targeted excitation methods. In quasi-resonant schemes (b) (ii), since the excitation energy is closer to the actual QD states, there are fewer relaxation steps involving phonons or other dephasing interactions, leading to a more coherent emission in comparison to the above band scheme \cite{Phillips2024,Dusanowski:17}. Finally, moving to the resonant scheme (b) (iii), the excitation photons have precisely the energy needed to excite the QD directly from the ground state to an excited state. By directly pumping a single QD transition, we minimise the instability associated with that transition, leading to enhanced photon indistinguishably and coherence, and minimum time jitter \cite{unsleber2016highly, H1Cav2_2018}. This makes it desirable for deterministic single-photon generation, which is essential for large-scale quantum networks \cite{Lodahl_network}. For the resonant measurements, the Tri-defect was excited from above with a resonant laser, scanning from 954.1nm to 954.25nm. The emission of the QD was modulated using an applied bias, to allow for the subtraction of the background laser scatter from the RF signal. To ensure a high signal-to-noise ratio, we introduced a cross-polarisation scheme that introduced a phase of $\pi/2$ between the linearly polarised pump laser and QD emission. A more detailed description of this process is discussed in the supplementary material S6. Fig \ref{fig:resonance} (c) compares the non-resonant emission from a single QD excited with a 808nm laser (red) with the equivalent quasi-resonant (P-shell, excited at 939.33 nm) and resonant scheme that delivers emission at 954.16nm. Here a significant reduction in linewidth to $8.9 \pm  0.4\mu eV$ (resonant) from $40.5 \pm  0.6\mu eV$ (non-resonant) is observed indicating a significant reduction in dephasing.
 
In Fig \ref{fig:resonance} (d) the normalised RF signal at different applied biases is shown. An RF peak from the emitter is present, and tunes 0.7nm from 954.15nm to 954.22nm as the bias is changed. As illustrated in Fig \ref{fig:resonance} (c,d), the absence of fine structure splitting in the emission spectrum suggests the QD is in a charged exciton state, either $X^{-}$ (negatively charged exciton) or $X^{+}$ (positively charged exciton). We believe the ability to observe stable charged states can be attributable to the combined charge stabilisation/state tuning from the p-i-n diode structure and the minimised charge noise by having the QD in the Tri-defect region. In order to gauge the single photon behaviour, we performed a Hanbury Brown and Twiss (HBT) auto-correlation measurement under a quasi-resonant excitation  giving us a second-order correlation function, $g^{(2)}(\tau)$ (as a function of coincidence time delay $\tau$). From Fig.\ref{fig:resonance} (e) at zero time delay ($\tau=0$) we have observed strong anti bunching such that,  $g^{(2)}(0) = 0.14 \pm 0.05$ showing we are operating in the single photon regime from a QD inside the Tri-defect region. More information on this measurement can be found in supplementary section S7.

\subsection{Resonant Transmission}

By moving into a resonant transmission scheme (seen in Fig \ref{fig:transmission} (a)), the nonlinear response of the QD within the Tri-defect region can be investigated. A QD that is well coupled to an optical mode behaves non linearly at the few-photon level. Single photons are reflected by the emitter, while multi-photon states are more likely to be transmitted. The observation of this phenomenon in this device indicates a strong interaction between the QD in the defect and photons in the topologically protected waveguide mode. Such phenomena underscore the potential of defect regions to enhance the interaction between photons mediated by topologically protected modes. 

In the data presented in Fig \ref{fig:transmission}, continuous-wave pumping at resonant wavelengths and voltages are used to probe these nonlinear behaviours. At low excitation powers, on the order of 10 nW, the strongly attenuated guided laser light predominantly occupies zero and one photon states within a single QD emission cycle. 

When the wavelength of the incoming photons is resonant with the QD transition, the system exhibits a characteristic dip in the transmission spectrum, indicative of photon reflection by the QD. Fig \ref{fig:transmission} (a) shows the waveguide transmission as a function of laser wavelength and applied voltage when the laser is tuned across the $X^{\pm}$ state of the QD. The transmission is normalized at each point to the transmission measured with the QD in an optically inactive state (at a bias of 1 V, in this case).

Fig \ref{fig:transmission} (c) shows a snapshot of (b) with the excitation laser fixed at 954.16 nm, where it is possible to see the strongest transmission dip with a minimal Fano lineshape ($\approx 8\%$ dip). The degree and width of the transmission dip allow us to model the cavity-waveguide coupling efficiency into the propagating modes. We can place a lower bound on this coupling from a fit to the transmission spectra for 50nW (by taking an upper bound on the resonant decay rate as the measured quasi-resonant decay rate shown in supplementary section S5), giving a coupling efficiency of 61 $\pm$ 4 $\%$ (details of calculation in supplementary section S9). In Fig \ref{fig:transmission} (d), at an excitation wavelength of 954.16 nm, as we increase the input laser power we observe a transition towards higher photon occupancy within the laser field, which diminishes the transmission minimum via dot saturation (from $\approx 7\%$ at 50nW to $\approx 0.5\%$ at 10 $\mu$W). The relationship between the transmission and the power is given by:
\begin{equation}
    T=1-A\left({1+\frac{P}{P_c}}\right)^{-1}
\end{equation}
Where $P_{c}$ is the critical power at which one photon within the time period of the QD's lifetime, couples through the QD, and A is the minimum transmission dip. A least squares fit (represented by the dashed line) to the experimental data presented in Fig \ref{fig:transmission} (d), gives $A$ = 0.071 and $P_{c}$ = 388nW. This power dependent transmission behaviour shows the nonlinear nature of the interaction.

By characterising the photon statistics of this nonlinear response, we can gauge the system's ability to sustain the highly coherent interactions of the resonant scattering (operating in the coherent scattering regime using a pump power of 50 nW). In an ideal case, the QD will reflect single photon components and only photon-photon bound pairs (multi-photon states) will transmit, leading to a bunching effect at $ \tau = 0$ in a  $g^{(2)}(\tau)$ measurement. In Fig \ref{fig:transmission} (e), we can see a distinct bunching peak of $g^{(2)}(0)=1.04$ indicating the coherent scattering of single photon components as predicted. This behaviour would be difficult to observe if not for a sufficiently high coupling efficiency from the cavity into the topological waveguide mode, and efficient collection of the light-by mode adapters. This shows promise that these devices can become integrated into larger scalable systems.

\section{Conclusion}

The successful demonstration of RF of QDs embedded within topological photonic crystal waveguides marks a notable advancement towards realizing on-chip integrated quantum optical devices. The integration of QDs within defects provides enhanced stability and efficiency for single-photon sources, crucial for quantum computing, communication, and sensing technologies. The use of an inverse designed connection between the topological waveguide and conventional nanobeam structures significantly improves the coupling efficiency, facilitating better integration into larger photonic circuits. Additionally, the enhancement in chiral coupling within the defect-engineered regions enables precise control over the emission directionality of single photons, which is a critical factor for advanced quantum communication protocols. The exploration of nonlinear optical effects at the single-photon level within such topologically protected environments addresses a previously ongoing challenge in the field. The ability to manipulate single photons within these systems, as evidenced by the observed resonant/nonlinear behaviour under different excitation conditions, opens up new avenues for research and application. The engineered defect regions within photonic crystal waveguides, facilitating these interactions, underscore the potential of topological photonics in enhancing light-matter interactions at the quantum level.

Looking forward, coupling multiple QDs to the edge states \cite{Grim2019} within topological photonic systems presents an exciting direction for further research. The collective dynamics between distant emitters \cite{science_lodahl}, within such a framework could unveil new quantum phenomena and enable the development of more complex and scalable quantum photonic circuits. The interplay between the quantum emitters and the topological modes, as facilitated by the engineered defects, offers a rich platform for exploring new regimes of light-matter interaction. This could lead to the realization of novel quantum optical devices and functionalities, such as topologically protected quantum gates or routers, which are essential for the advancement of quantum information technologies.


\section*{Funding}
This work was supported by EPSRC Grant No. EP/N031776/1 and EP/V026496/1 and the Quantum Communications Hub EP/T001011/1. 

\section*{Acknowledgments}
The authors would like to acknowledge helpful discussions with Mohammad Hafezi, Peter Millington-Hotze and Catherine Phillips.
\\
\section*{Author contributions statement}
N.J.M., M.J.M., designed the photonic structures, which R.D. fabricated. E.C. and P.K.P. grew the sample. N.J.M., X.C., L.H. carried out the simulations of the devices. N.J.M., L.H., D.H., L.B. and A.F. carried out the measurements. L.R.W., A.M.F, M.H., and M.S.S. provided supervision and expertise. L.H., N.J.M., M.J.M., wrote the manuscript, with input from all authors. L.H., N.J.M contributed equally to this work. 

\section*{Data availability} Data underlying the results presented in this paper are not publicly available at this time but may be obtained from the authors upon reasonable request.

\section*{Supplemental document}
See Supplement 1 for supporting content.

\bibliography{Bibli}

\widetext
\clearpage

\begin{center}
\textbf{\large Supplemental Materials: Nonlinear Quantum Optics at a Topological Interface Enabled by Defect Engineering}
\end{center}
\newpage
\setcounter{equation}{0}
\setcounter{figure}{0}
\setcounter{table}{0}
\setcounter{page}{1}
\setcounter{section}{0}
\makeatletter
\renewcommand{\theequation}{S\arabic{equation}}
\renewcommand{\thefigure}{S\arabic{figure}}

\renewcommand{\thesection}{S\arabic{section}}

\section{Inverse Design of connection \label{SI:INVERSE}}

\begin{figure}[h]
    \centering
    \includegraphics[width=0.65\textwidth]{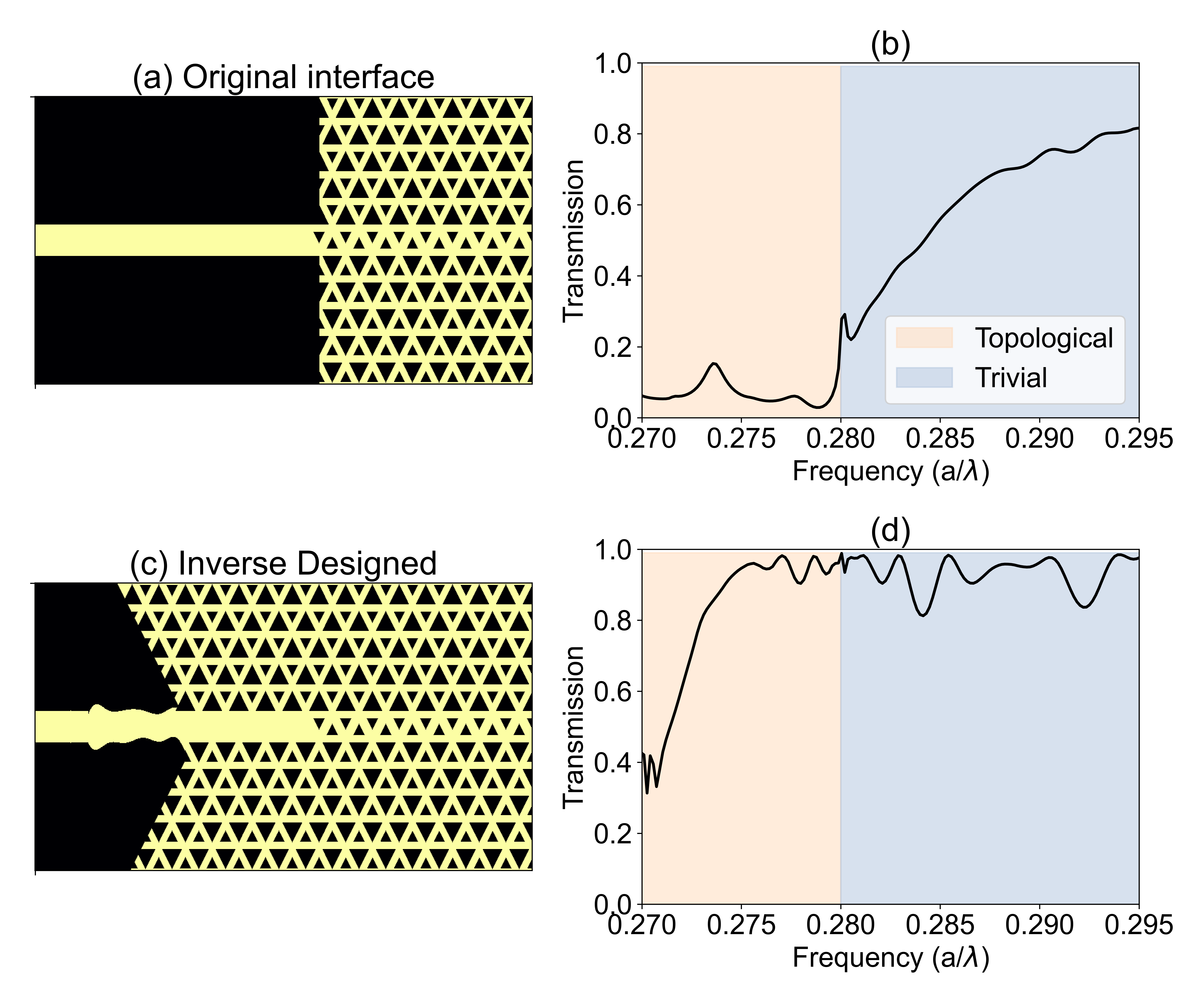}
    \caption{Design of an un-optimised interface (a), and optimised interface (c) with their respective transmission properties in (b) and (d).  }
    \label{fig:con_inverse_design}
\end{figure}

For VH waveguides and devices to be successfully integrated into larger integrated optics geometries, their efficient coupling to simple slab waveguides is essential. Previous efforts to create an enhanced interface have focused on decreasing the mode waist radius \cite{tapering} and the introduction of a line defect region \cite{Yoshimi:24}, however these methods don't take into account the change in the structural symmetry from the nanobeam waveguide to the photonic crystal. The `bearded' interface VH waveguide, used in this work, has glide symmetry at the interface, whilst the nanobeam has mirror symmetry about the x-axis. Any improved connection must include a taper from one to the other, such as those used for glide plane waveguides \cite{hamidreza, lodahl_glide_plane}. In order to create such a taper, we employed an inverse design technique. Using the adjoint based gradient descent method of a commercial Maxwell solver \cite{Lumerical}, where two FDTD simulations are used to compute the gradient of the figure of merit (a forward simulation and an adjoint), we were able to optimise the transmission by optimising the shape of a multi parameter taper. The speed of the gradient decent method makes this an extremely efficient technique for the design of complex, multi-parameter, integrated photonics components  such as this \cite{Lalau-Keraly:13}. 

The figure of merit for the optimization was the forward transmission through the waveguides, measured as the transmission into a selected modal field rather than the total transmitted power. This approach ensures that the optimization targets genuine transmission through the nanobeam, avoiding solutions that maximize transmission by increasing random scattering captured by a power monitor. The optimization process involved 20 parameters, all optimized simultaneously, a significantly higher number than typically feasible with standard particle swarm optimization.

The result of this optimization is depicted in Fig. \ref{fig:con_inverse_design}. Fig \ref{fig:con_inverse_design} (b) and (d) show a simulated comparison of the transmission for an un-optimised (a) and the optimised (c) connection. The transmission plots simulated using 2.5FDTD (or VarFDTD) show the topologically trivial and non-trivial operation of the waveguide, with the slow light region of operation being around 0.275 $a/\lambda$, where we see a significant improvement of the transmission. Without the taper, less than 10\% of light in the topologically non-trivial waveguide mode couples to the nanobeam. This taper improves the performance such that the transmission is $>90\%$ over a wide spectral range.



\newpage

\section{Chiral Coupling Within Tri Defect}

\begin{figure}[h]
    \centering
    \includegraphics[width=0.7\textwidth]{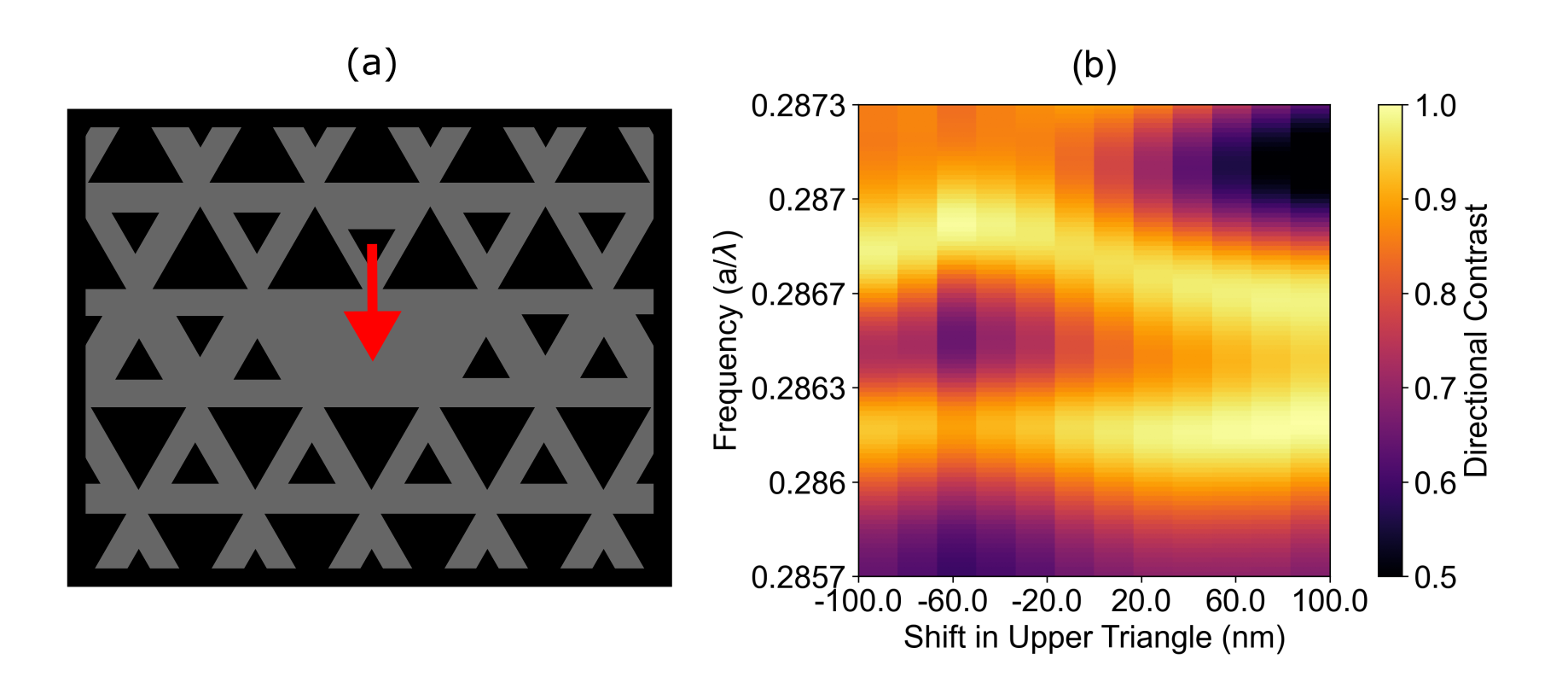}
    \caption{(a) Pictorial representation of the movement of one of the triangles around the Tri-defect in FDTD simulations of the waveguide coupled defect cavity. (b) Plot of the simulated wavelength dependence for the directional contrast of a circularly polarised dipole emitter at the the centre of the defect, for different shifts in the upper triangle.  }
    \label{fig:chiral_tri}
\end{figure}

In the Tri-defect, the directionality arises from a interference effect between its two cavity modes \cite{ChiralCav_2022}. A circular dipole will excite a superposition of these two cavity modes with a known phase difference ($\pm \pi$/2, depending on the dipole handedness). With the addition of an phase difference arising from the detuning between the emitter and each cavity mode, if the amplitudes of the fields are the same, complete destructive interference can be achieved in one waveguide, and constructive in the other, and directional emission is realized. Figure \ref{fig:chiral_tri} shows how the detuning between the modes can be arbitrarily changed, to form different directional contrasts

\newpage
\section{$\beta$-Factor and Cavity-Waveguide Coupling}

\begin{figure}[h]
    \centering
    \includegraphics[width=0.7\textwidth]{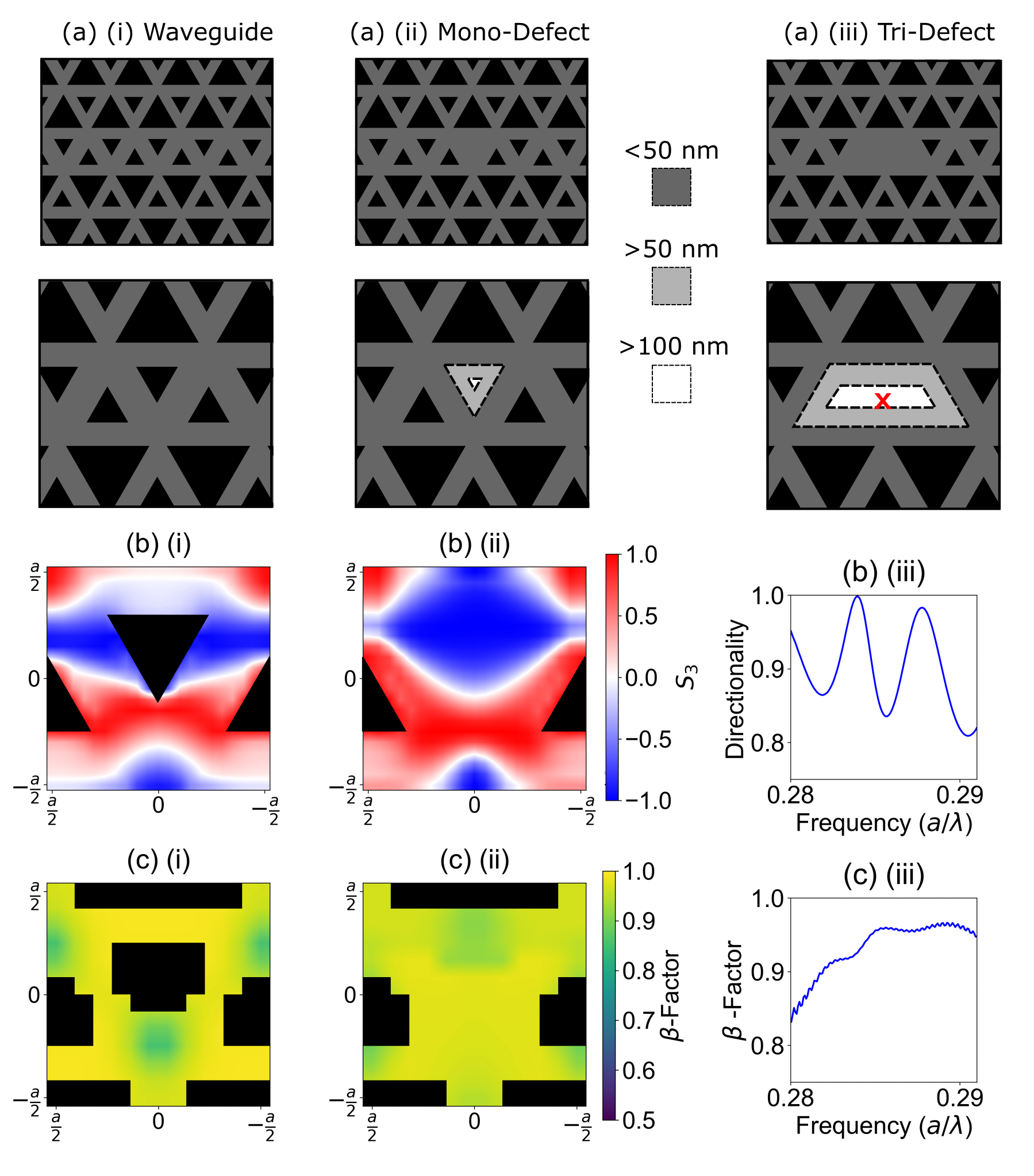}
    \caption{Schematic of the (a) (i) unmodified waveguide structure, (a) (ii) Mon-defect structure and (a) (iii) Tri-defect structure. Figure (b) (i-ii) shows the stokes $S_{3}$ parameter, characterising the degree of circular polarisation of the electric field at the centre of the waveguide slab. (b) (iii) shows the equivalent wavelength dependence of the directionality for the Tri-Defect. (c) (i-ii) shows the $\beta$-factor calculated using 3D FDTD simulations for the unmodified waveguide (i) and the mono-defect (ii), with the equivalent wavelength dependent coupling efficiency for the tri-defect. }
    \label{fig:enter-label}
\end{figure}

\newpage

\newpage

\section{Wafer and Electrical Properties \label{SI:Wafer}}

\begin{figure}[h]
    \centering
    \includegraphics[width=0.8\textwidth]{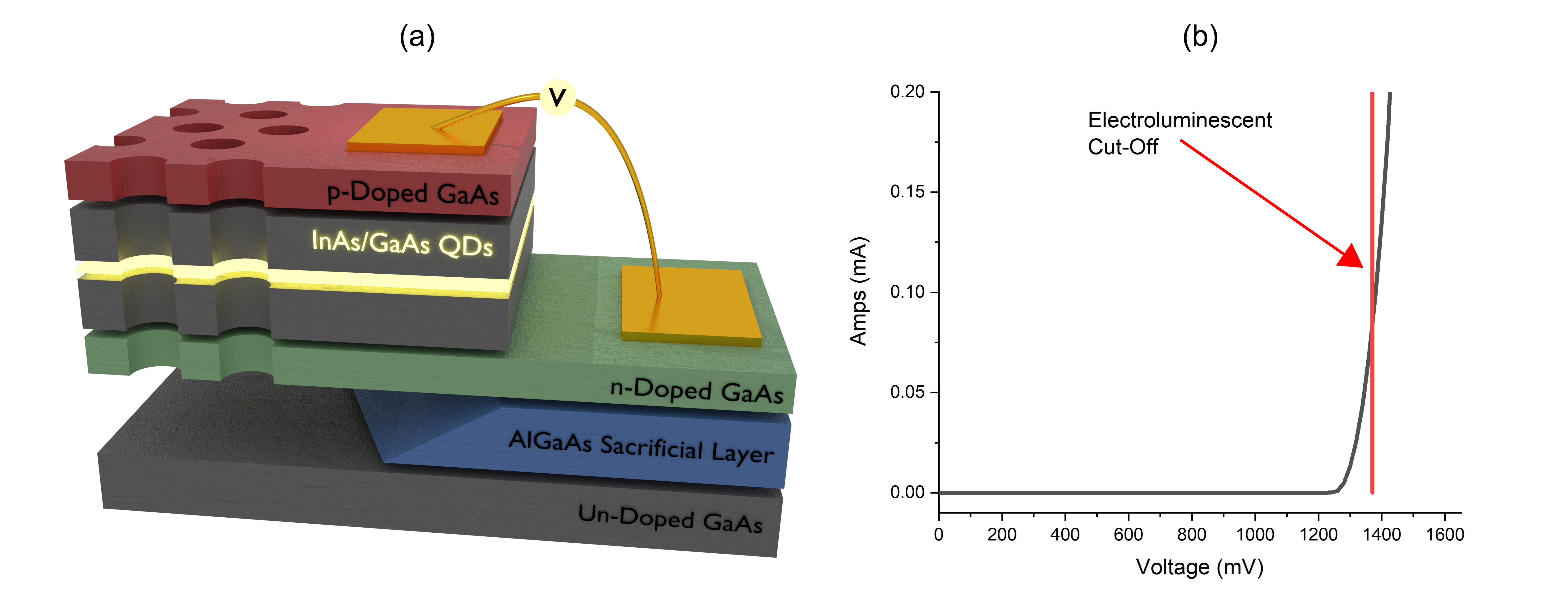}
    \caption{a) Schematic of p-i-n diode structure used for topological waveguide structures. With gold contacts to apply bias across the undoped region. b) IV curve of sample diode with inflection point of electroluminescence indicated by vertical red line}
    \label{fig:wafer}
\end{figure}

In the centre of the intrinsic region lies a 0.7nm layer of InAs quantum QD (QDs) formed through strain relaxation, caused by a 17\% lattice mismatch with undoped GaAs (referred to as Stranski-Krastanov QD) emitting between 910nm to 980nm. The membrane structure (170 nm thick) is grown on a 1$\mu$m thick AlGaAs sacrificial layer using MBE (molecular beam epitaxy). Using this material, we can easily under-etch the membrane structure using wet-etching with HF solution to ensure maximum optical confinement within the topological structure via air cladding. In Fig.\ref{fig:wafer}b, the I-V curve demonstrates the typical diode-like response to an applied bias across the p and n layers (connected via gold contacts). At the inflexion point indicated by the solid red line in the I-V graph, electroluminescence becomes the dominant excitation mechanism of the QDs.

\newpage
\section{Lifetime \label{SI:LIFETIME}}
\begin{figure}[h]
    \centering
    \includegraphics[width=0.6\textwidth]{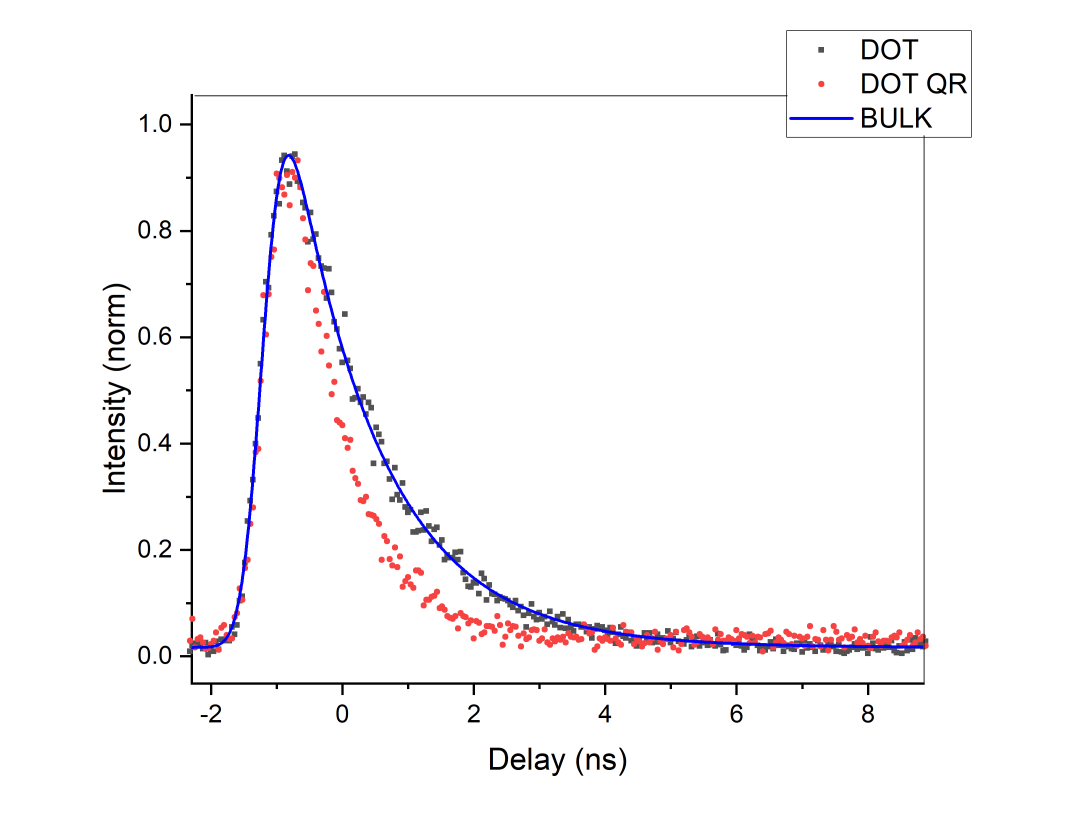}
    \caption{Time correlated measurement showing excitation and recombination rate of QD under non resonant (NR) and quasi resonant (QR) pumping schemes}
    \label{fig:enter-label}
\end{figure}

In our investigation of this QD we also measured the radiative lifetime (fluorescent lifetime) using Time correlated single photon counting techniques to measure the populations of charge carriers over a sufficient time period to minimise Poisson noise. This was acheived using a femtosecond Tsunami Ultrafast Ti:Sapphire pulsing laser with time correlated single photon counting on a Swabian Instruments time tagger Ultra. The experimental data was fitted using a exponential decay of the form $e^{\frac{-\tau}{\gamma}}$ were gamma represents the fluorescent lifetime.
\\
\\
We observe no large change in emission rate compared to bulk lifetimes. However under a quasi resonant scheme we observe a radiative lifetime of 0.89 ± 0.02 ns compared  to 1.38 ± 0.01 ns in the non-resonant scheme. 
\\
\\

\newpage
\section{Resonant Measurement Setup\label{SI:SETUP}}
\begin{figure}[h]
    \centering
    \includegraphics[width=0.7\textwidth]{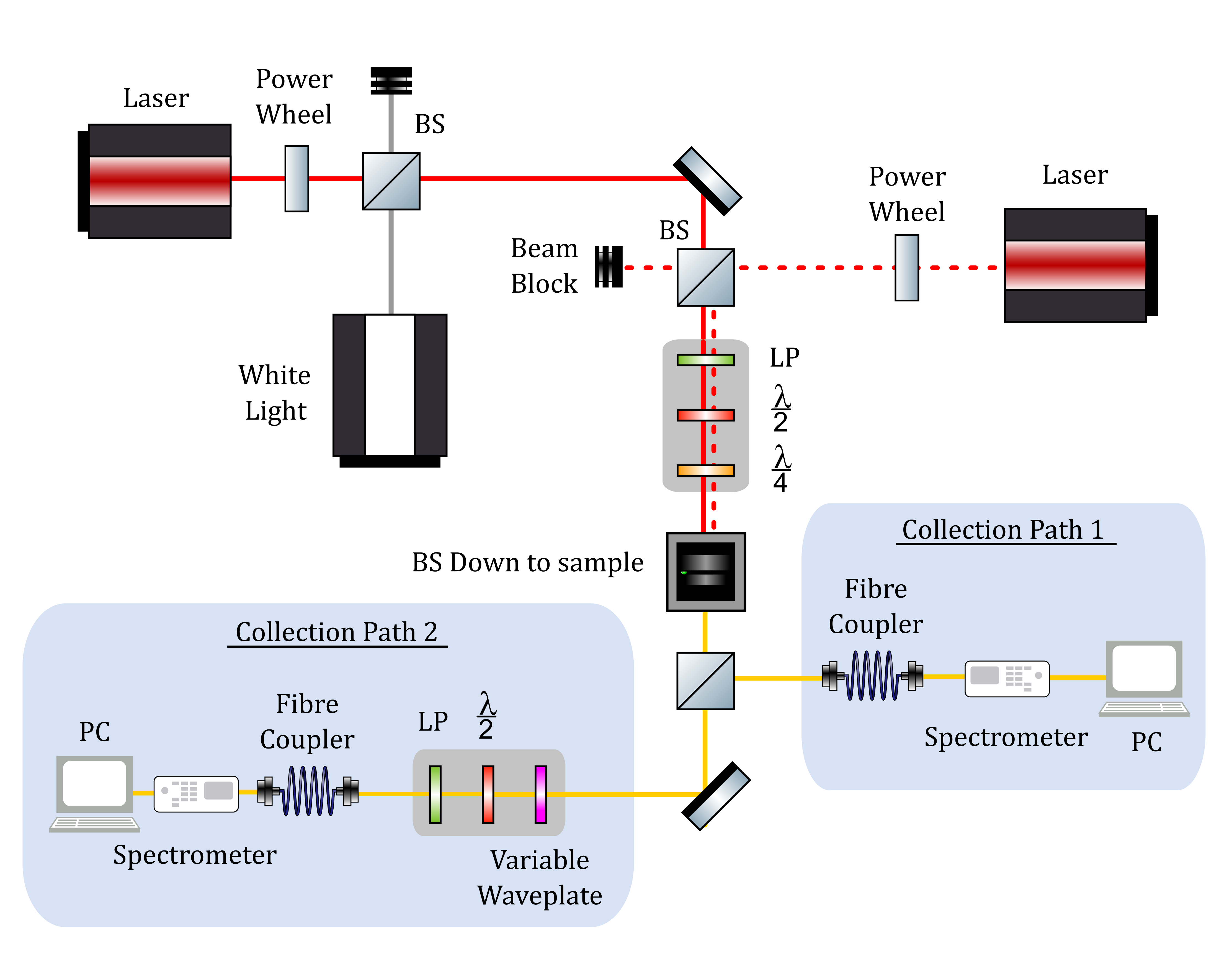}
    \caption{Schematic of Micro-PL setup. Cross polerisation RF acheived by  placing wave plates on the excitation to control linear and circular polarisation and achieve a $\frac{\pi}{2}$ phase between the  }
    \label{fig:enter-label}
\end{figure}

Fig. S7 illustrates the resonance fluorescence scheme used to acquire the data in the main text. The laser and dot emission collected in the far-field scattering from an out coupler are linearly polarised. We can take advantage of this to modify the polarisation between the excitation and collection path using a series of linear polarisers (LP), half waveplates ($\frac{\lambda}{2}$) and quarter waveplates ($\frac{\lambda}{4}$)/adjustable waveplates to ensure a $\frac{\pi}{2}$ difference in polarisation angle. This results in the scattered laser light being rejected en route to the spectrometer or APD. Voltage modulation allows us to filter out background fluctuations caused by electroluminescence or scattering processes from the pump laser during measurements and compensate for changes in laser power resulting from tuning/instability.

\newpage

\section{Quasi-Resonant Cross Correlation Measurement\label{SI:SETUP}}

The second-order correlation function \(g^{(2)}(\tau)\) for the Quasi-Resonant excitation of a QD can be modeled as \cite{PhysRevB.104.075301,PhysRevB.92.245439}:

\begin{equation}
g^{(2)}(\tau) = \left[\left(g_0 - 1\right) e^{-\frac{|\tau |}{T_{\text{corr}}}} + 1\right] 
\left[\left(\frac{1}{\beta} - 1\right) e^{-\frac{|\tau |}{T_{\text{blink}}}} + 1\right]
\end{equation}

where:
\begin{itemize}
    \item \(T_{\text{corr}}\) is the characteristic correlation timescale.
    \item \(T_{\text{blink}}\) is the characteristic blinking timescale.
    \item \(\beta\) is the blinking on-off ratio.
    \item \(g_0\) is the $g^{(2)}(\tau)$ value at $(\tau)$=0 .
\end{itemize}

\section*{Fitted Parameters}

The fitted parameters and their corresponding errors are presented in the table below:

\begin{table}[h!]
\centering
\begin{tabular}{|l|c|c|}
\hline
\textbf{Parameter} & \textbf{Value} & \textbf{Error} \\
\hline
\(T_{\text{corr}}\) (ns) & 0.75 & $\pm$ 0.06 \\
\(T_{\text{blink}}\) (ns) & 49 & $\pm$ 1 \\
\(\beta\) & 0.547 & $\pm$ 0.004 \\
\(g_0\) & 0.14 & $\pm$ 0.05 \\
\hline
\end{tabular}
\label{tab:fit_parameters}
\end{table}

\newpage
\section{Transmission Modelling Factor\label{SI:SETUP}}
To model we used to fit the transmission dip in Fig 4 was first presented in \cite{hallett2018electrical}, which we repeat here for clarity: 
\subsection{Input-Output Relations}

We consider a system where a QD is coupled to a nanophotonic waveguide. Neglecting dissipative dynamics, the Hamiltonian $H = H_0 + H_{int}$ is given by \cite{PhysRevA.82.063821} :

\begin{align}  
H_0 &= \hbar \int d\epsilon (\omega_0 + \epsilon) (r_\epsilon^{\dagger} r_\epsilon + l_\epsilon^{\dagger} l_\epsilon)
\\
H_{int} &= \frac{1}{2} \hbar \Omega \sigma_z + \hbar \int d\epsilon \left[(g_r r_\epsilon \sigma^+ + g_l l_\epsilon \sigma^-) + \text{H.c.}\right]
\end{align}

Here, $r_\epsilon$ and $l_\epsilon$ are the annihilation operators for right- and left-propagating photons with frequency $\omega_0 + \epsilon$. The frequency of the $\ket{e} \rightarrow \ket{g}$ transition is $\Omega$, and $g_r$, $g_l$ are the coupling amplitudes. $\sigma^+$ and $\sigma^-$ are the Pauli operators.

\subsection{Dephasing and Coupling Into Unguided Modes}

Using the Input-Output formalism of \cite{PhysRevA.30.1386}, we derive the coupled differential equations for the left and right input fields and QD dynamics. Following the work in \cite{PhysRevA.75.053823}, we add to the dynamics a finite pure dephasing and spontaneous emission into unguided modes. Where dephasing time is represented by $\tau_d$, spontaneous emission rate is given by $\gamma'$, and defining $\beta$ as the fraction of emission into guided modes and $\beta_d$ as the fraction into right-propagating modes, we find:

\begin{align}
r_{\text{out}} &= r_{\text{in}} - i\sqrt{\frac{\beta_d \beta}{\tau}} \sigma_- \quad,\quad 
l_{\text{out}} = l_{\text{in}} - i\sqrt{\frac{(1 - \beta_d) \beta}{\tau}} \sigma_- \label{in_out}
\\
\dot{\sigma}_- &= - \left(i\Omega + \frac{1}{\tau} \right) \sigma_- + i \sigma_z \left[ \sqrt{\frac{\beta_d \beta}{\tau}} r_{\text{in}} + \sqrt{\frac{(1 - \beta_d) \beta}{\tau}} l_{\text{in}} \right]
\label{sigma}
\\
\dot{N} &= - \frac{1}{\tau} N + i \left[ \sqrt{\frac{\beta_d \beta}{\tau}} (r_{\text{in}} \sigma_- - r_{\text{in}}^\dagger \sigma_+) \right] + i \left[ \sqrt{\frac{(1 - \beta_d) \beta}{\tau}} (l_{\text{in}} \sigma_- - l_{\text{in}}^\dagger \sigma_+) \right]
\end{align}

where $2\gamma = 2\gamma_a + \gamma'/2 + 2\pi g_r^2 + 2\pi g_l^2$ and $N = (\sigma_z + 1)/2$ gives the emitter population. 

The total emitter lifetime $\tau$ and the $\beta$-factor are modified to be: 

\begin{align}
\tau \rightarrow \tau' &= \tau / [F_p \beta + (1 - \beta)]
\\
\beta \rightarrow \beta' &= F_p \beta / [F_p \beta + (1 - \beta)]
\end{align}

In the weak excitation regime, replace $\sigma_-$ in Eq. \ref{sigma} by -1. For coherent inputs $r_{\text{in}}$ and $l_{\text{in}}$, integrate Eq. \ref{sigma} and substitute into Eq. \ref{in_out} to find the output fields. For a QD driven from the left with amplitude $r$ and frequency $\Omega$, the transmission spectrum is:

\begin{equation}
T(\omega) = \left|\frac{\langle r_{\text{out}}(\omega)\rangle}{\langle r_{\text{in}}(\omega)\rangle}\right|^2    
\end{equation}

\subsection{Spectral Diffusion and Blinking}

Charge noise causes spectral wandering of the exciton energy, characterized by variance $\sigma$. The transmitted intensity $T(\Omega)$ must be modified:

\begin{equation}
T(\Omega) \rightarrow T(\Omega, \sigma) = \frac{1}{\sqrt{2\pi\sigma^2}} \int d\epsilon e^{-\epsilon^2 / 2\sigma^2} T(\Omega + \epsilon)
\end{equation}

Considering the finite probability $P_{\text{dark}}$ of the QD being in an inactive `dark' state, modify as:

\begin{equation}
T(\Omega) \rightarrow T(\Omega, \sigma, P_{\text{dark}}) = (1 - P_{\text{dark}}) T(\Omega, \sigma) + P_{\text{dark}}
\end{equation}





\subsection{Parameters used for the Transfer Matrix Model}
For the fit presented in Fig 4(c), the modification of the transmission due to the presence of dark states was omitted by setting the probability $P_{dark}=0$. The $\beta-$ factor of 0.61 $\pm$ 0.04, can be considered a lower bound to the true value. 
\begin{table}[h]
    \centering
    \begin{tabular}{|c|c|c|} \hline 
         \textbf{Parameter}&  \textbf{Value}& \\ \hline 
         Lifetime&  890 ps& Measured\\ \hline 
         Central Wavelength&  954.16 nm& Measured\\ \hline 
         Pure dephasing time&  74 ps& Measured\\ \hline 
         Variance of
spectral
wandering&  0.74 $\mu$eV& Fitting Parameter\\ \hline 
         $\beta$-Factor&  0.62& Fitting Parameter\\ \hline 
         Non-Resonant Transmission&  0.926& Fitting Parameter\\ \hline
    \end{tabular}
    
    \label{tab:my_label}
\end{table}

\end{document}